\documentclass[fleqn,twoside]{article}
\usepackage{amsmath}
\usepackage{espcrc2}

\usepackage{graphicx}

\def\p{\partial }
\def \cO{{\cal O}}
\def\e{\epsilon}
\def\beq{\begin{equation}}
\def\eeq{\end{equation}}
\def\ba{\beq\begin{array}{c}}
\def\ea{\end{array}\eeq}
\def\be{\ba}
\def\ee{\ea}

\def\p{\partial}

\def\g{{ \gamma}}

\def\e{{\epsilon}}

\title{ Carg\`ese Lectures on the Kerr/CFT Correspondence}

\author{Irene Bredberg, Cynthia Keeler, Vyacheslav Lysov  and Andrew Strominger \\ \hspace{20 mm} \\Center for the Fundamental Laws of Nature\\ Harvard University\\ Cambridge, MA 02138 USA}

\begin{document}

\begin{abstract}
We give a short introduction, beginning with the Kerr geometry itself, to the basic results, motivation,  open problems and future directions of the Kerr/CFT correspondence. \vspace{1pc}
\end{abstract}

\maketitle

\tableofcontents

\section{Introduction}

In the early 1970's, work by  Bekenstein, Carter, Christodolou, Hawking, and many others \cite{Christodoulou:1970wf,Hawking:1971tu,Bardeen:1973gs,Bekenstein:1973ur,Bekenstein:1973mi,Bekenstein:1974ax,Hawking:1974sw} raised profound puzzles about the nature of black holes. One striking such puzzle was that, while macroscopic arguments gave the
entropy of a black hole as one quarter of its event horizon area:
\begin{equation}\label{entr}
~~~~~~~~~~~~~~~~~~~~~~S = {A\over 4\hbar G},
\end{equation}
at the time no microscopic accounting for this entropy was known.
It seemed imperative that we should be able to account for the black hole entropy microscopically, just as had been done in the nineteenth century
 for gases and liquids. Without such a microstate description, we would seem to run into serious contradictions.

This problem remained largely unsolved
for more than 20 years. Then in the mid 90's string theory was used \cite{StromingerSH} to explicitly identify the missing microscopic degrees of freedom for a very particular kind of black hole. This calculation  depended on many specific details of string theory. At the end of a rather lengthy computation involving numerous factors of 2, $\pi$ etc., the Bekenstein-Hawking result
(\ref{entr})  was reproduced by counting microstates.  At the time, it was argued that this precise match provided indirect evidence for string theory as the correct theory of nature.

However, about a year later, it was shown \cite{StromingerEQ} that in fact, any consistent, unitary quantum theory of gravity containing those particular black holes - characterized by a near-horizon region with an $AdS_3$ factor - as solutions
must reproduce the entropy in essentially the same way. The specific details of string theory as the microscopic UV completion were not necessary. Rather, the key ingredient followed from the analysis done by Brown and Henneaux \cite{BrownNW} in the 80's: if we find a consistent completion of quantum gravity on $AdS_3$ it has to be described by a 2D  conformal field theory due to purely symmetry considerations.  Thus, the detailed matching of the factors of 2 and $\pi$ was not really a consequence of string theory but rather, it simply had to follow because  string theory is a consistent theory of quantum gravity.  Any other consistent theory must by necessity also reproduce the same result in the same manner.\footnote{The other side of the coin here is that these general arguments  imply that
any consistent quantum theory of gravity must, on an AdS$_3$ background, behave a lot like string theory -- so much so that we might reasonably call it string theory!}

Since then, we have slowly but surely been progressing in our understanding of the relation between black holes and 2D CFTs. We started with 5D  supersymmetric black holes,
then proceeded to partially supersymmetric and then to the 3D  nonsupersymmetric black holes with near-horizon $AdS_3$ geometry. Recently, our understanding has finally evolved to up the point where we can understand something  about 4D  Kerr black holes  that we see up in the sky.

The work we are going to discuss is heavily informed by string theory, but none of it relies on the conjecture that string theory is the actual theory of nature, or on the stringy realizations of the AdS/CFT correspondence. Instead, all of our arguments follow from careful study of the diffeomorphism group together with some basic consistency assumptions, and do not involve any details of Planck scale physics. Indeed it would be very strange if the universal area-entropy law somehow depended on the exact microscopic details of how quantum gravity is completed in the UV!

To emphasize this point further, let us draw an analogy of the current efforts with the work of Boltzmann in the 19th century. At that time thermodynamics was understood, but people did
not know much about atoms and molecules. Boltzmann wanted to explain the laws of thermodynamics by applying statistical, probabilistic
reasoning to the fundamental constituents (degrees of freedom) of gases and liquids.  However, he encountered a UV divergence: if a gas is treated as
a continuous medium, then it has infinitely many degrees of freedom because of the existence of arbitrarily short wavelength modes. Any attempt to derive the thermodynamics of gases by applying statistical reasoning
to a theory of a continuous medium, will hit the so called Rayleigh ultraviolet catastrophe in which all energy is eventually sucked into the UV modes. To avoid this problem,  a consistent UV cutoff is needed. People were already talking at that time about atoms and molecules, so Boltzmann assumed that there was some theory of atoms, i.e. he assumed that there was a consistent UV cutoff for gases and liquids. He did not at all need to know what the details of this atomic cutoff were; in fact, the periodic table was not discovered until more than fifty years later. Boltzmann's mere assumption that there existed a UV cutoff at some energy scale was sufficient to derive the universal laws of thermodynamics from statistical reasoning.  Of course, having a detailed microscopic theory can provide more information; for example if one wants to compute the heat capacity from first principles, one needs a detailed UV completion (that is, the actual quantum theory of atoms and molecules).

We might hope that a similar story holds for black holes. We should not need to know all the details of string theory at scales of order $10^{-38}$ km in order to understand why the area law (\ref{entr}) applies to the black hole Sagittarius A* in the center of our galaxy which is $10^7$ km across! We should be able to understand the area law just from the assumption that quantum gravity has {\it some} consistent UV completion.

The stringy microscopic entropy analysis in  \cite{StromingerSH} was akin to first computing the periodic table and then using it to compute the laws of thermodynamics. In this stringy black hole computation we had far more information than was necessary to get the area law: we had huge sets of numbers for degeneracies at any level. Only a tiny part of this information turns out to be universal. We are going to see in these lectures that this  tiny universal part can be understood using universal reasoning and no assumptions about Planck scale cutoffs. This is exactly as it should be.

In these lectures we will encounter another much-studied object in theoretical physics which has a lot of universal behavior: 2D conformal field theories. Many features we know of 2D CFTs are independent of the details of a given CFT. Indeed, we will find a striking  match  -going far beyond the entropy formula (\ref{entr}) - between the universal properties of 2D conformal field theories and those of black holes.

The plan for the rest of the lectures is the following: we will start with a review of the Kerr geometry, including the Near-Horizon Extreme Kerr (NHEK) geometry. Then we will cover the asymptotic symmetry group, boundary conditions for the NHEK geometry, the CFT description of a quantum theory of gravity in NHEK  and the surprising evidence for hidden conformal symmetries far from extremality. We will close with a discussion of open problems and future directions.

\section{Kerr geometry}
\subsection{The Kerr solution}
There is a famous quote from Chandrasekhar \cite{Chandrasekhar}

\noindent ``.... {\it Kerr's solution has also surpassing theoretical
interest: it has many properties that have the aura of the miraculous about them}. ''

\noindent The first ``miracle'' of the Kerr story is the existence of the solution itself.
It is rumored that  Einstein initially  believed that no non-trivial exact solution to the GR equations would ever be found in closed form. This was quickly proven wrong by Schwarzschild, but it took another 50 years to discover the Kerr solution. This solution is arguably the most complicated exact solution ever found of a nonlinear PDE  describing a real physical object. In Boyer-Lindquist coordinates,  it is
 \begin{multline} \label{kerr} ds^2 = - {\Delta \over \rho^2}\left(dt - a\sin^2\theta \, d\phi\right)^2 +{\rho^2\over \Delta} dr^2+\\
 {\sin^2\theta \over \rho^2}\left((r^2+a^2)d\phi -a\, dt\right)^2+\rho^2 \, d\theta^2,
 \end{multline}
 \beq\label{rhdl}
\Delta = r^2 -2Mr +a^2,\quad \rho^2 = r^2 + a^2 \cos^2 \theta.\eeq
with $a = J/M$ where $M$ is the mass and $J$ is the angular momentum of the black hole.\footnote{Here we have used units in which $G=c=1$.}
  The inner ($r_-$) and outer ($r_+$) horizons
 are defined as the two solutions to $\Delta =0$:
\beq\label{horiz} r_{\pm} = M\pm \sqrt{M^2-a^2}.\eeq
 When $a=0$, $r_+=2M$ and we recover the usual Schwarzschild black hole. Another interesting case is the extreme limit for which $a=M=\sqrt{J}$ and $r_{\pm}=M$. When  $J > M^2$ the roots in the expression for the horizon radii become imaginary; there are no horizons but instead a naked ring singularity in the curvature at $r=0$.  This violates cosmic censorship.  So cosmic censorship implies the angular momentum is bounded by $J\leq M^2$.
 Another way to understand this bound is from the formula for the angular velocity at the horizon:
 \beq\label{angm} \Omega_H = {a \over 2Mr_+}.\eeq
 When $a=M$, the equator of the horizon is spinning at the speed of light, and so cannot spin any faster.

 According to Bekenstein and Hawking, the black hole has a temperature and an entropy  \beq\label{tentr} T_H = {r_+-M \over 4\pi M r_+},\quad S = 2\pi Mr_+.\eeq
 Note that if $a=M$ then $T_H=0$, so these extreme rotating black holes are a kind of ground state. In general, the ground states of a system are easier to understand than
the excited states. The fact that $T_H=0$ in this special case hints that  $a=M$, rather than the Schwarzschild case $a=0$,  describes the simplest object in the Kerr family.
\subsection{Ergosphere}
The Kerr geometry has an interesting ergosphere region absent in the  Schwarzschild case.
Lines of fixed $\theta,\phi,t$ and varying $r$ in the Kerr geometry are space-like as long as we are outside the event horizon $r=r_+$. If we go to $r_-<r<r_+$,  then these lines become timelike,
just like in the Schwarzschild case when we cross $r=2M$.  For Schwarzschild,  lines with fixed $r,\theta,\phi$ and varying $t $ change from spacelike to timelike at the same surface $r=2M$.  However, for Kerr, these lines  switch signature at the stationary limit surface, which is given by the zeros of $g_{tt}$ and is outside of the event horizon.  This surface is described by solutions to  $\Delta = a^2 \sin^2\theta$ and is not spherically symmetric. The region in between $r=2M$ and the stationary limit surface is called the ergosphere.

Objects in the  Kerr geometry experience a ``frame-dragging'' force pushing them around the black hole. In order to stay at fixed $\phi$ outside a Kerr black hole we actually
  need to move with some speed in the counterrotational direction. This speed increases as we approach the stationary limit surface, and becomes the velocity of light when we reach it.
Once inside the stationary limit surface if we try to keep $\phi$ fixed while increasing  $t$, we end up moving in a space-like direction. Physically it means that inside this limit surface - in the ergosphere - physical objects cannot stay at constant $\phi$ because they would need to move with a speed greater than light to do so.  Instead, all timelike paths rotate in the same direction as the Kerr black hole.

  The ergosphere is a fascinating and  observable region of space-time where effects of general relativity are large.  Measurements which probe  the physics well inside the
  ergosphere near the event horizon  have been made (see \cite{McClintock:2011zq} for review). Perhaps the most-studied example is GRS1915+105 \cite{McClintock:2006xd}, a Kerr black hole surrounded by an accretion disc made up of matter which is continuously pulled in from a nearby companion star. As the matter in the accretion disc spirals inwards, energy is dissipated into thermal radiation which we observe.  The frequency distribution depends on a number of factors including the gravitational redshift which gets larger closer to the horizon.
The accretion disc has an innermost edge which corresponds to the existence of the so-called innermost stable circular orbit (ISCO) for geodesics in Kerr. Once matter gets closer to the black hole than the ISCO, orbits become unstable
  and the matter consequently free falls into the black hole rapidly and does not radiate anymore. Thus, the observed frequency spectrum encodes the location of the ISCO.  This location depends on the angular momentum of the black hole and is given by a standard GR calculation. For
  rapidly rotating black holes, the ISCO almost coincides with the event horizon, and the maximal redshift becomes very large. Hence the maximal redshift is very sensitive to the deviation from extremality. The observation of the ISCO thus allows us to get very accurate information about $J/M^2$ for
  black holes in the sky.  For example the lower bound on the ratio $J/M^2$ for $GRS 1915+105$ is $0.98$.   Its mass is known to much less precision; it is estimated to be between $5$ and $15$ solar masses.   The sky contains other  candidates for rapidly rotating black holes.  For example a measurement of the maximal redshift of the iron line for the supermassive black hole in the center of the nearby Seyfert galaxy  MCG-6-30-15 indicates $J/M^2$ is greater than $0.99$ \cite{Brenneman:2006hw}.  Given the rapid progression of our knowledge of the sky, it is quite possible  that more such near-extreme Kerr black holes will be discovered in the near future.
\subsection{Killing-Yano tensor}
 Another amazing property of  the Kerr geometry is the existence of a Killing-Yano tensor \cite{Walker:1970un}. It is an antisymmetric tensor with properties similar to those of Killing vectors. For a Killing vector $K_\mu$, we have
 \beq\label{kilng}  \nabla_{\mu}K_{\nu} + \nabla_{\nu}K_{\mu}=0.\eeq
The Killing-Yano tensor satisfies
 \beq\label{kyano} f_{\mu\nu} = -f_{\nu\mu},\quad \nabla_{(\lambda}f_{\mu)\nu}=0. \eeq
The explicit expression in Boyer-Lindquist coordinates is
\begin{multline}\label{eky} f = a\cos \theta dr\wedge (dt-a\sin^2\theta\, d\phi)\\
-r \sin\theta\, d\theta\wedge \left(-a dt +\left(r^2+a^2\right)d\phi\right).\end{multline}
This is yet another ``miracle''  of the Kerr story, since there is no obvious  reason for this tensor to exist. Perhaps we will learn something about it from Kerr/CFT, whose consistency relies on the existence of  this tensor.  Carter used (\ref{eky}) to construct a symmetric tensor \cite{Carte}
\beq\label{kkya} K_{\mu\nu} = f^{\phantom{\mu}\lambda}_\mu  f_{\lambda \nu},\;\; \nabla_{(\mu}K_{\nu\rho)}=0\eeq
and an extra conserved charge for geodesics:
\beq\label{cchar} Q= \dot{x}^\mu \dot{x}^\nu K_{\mu\nu} .\eeq
Consequently we have 3 constants of motion and can solve analytically for Kerr geodesics. It also allows us to
separate variables for the Laplace equation in the Kerr geometry.

Despite having an explicit expression for the  Killing-Yano tensor, we still do not fully understand its geometric origin. An intriguing  observation  has been made by Gibbons et. al. \cite{GibbonsAP}
relating it to a novel type of supersymmetry. The conserved Carter constant is the square of a novel supercharge (roughly $\psi^{\,\mu} f_{\mu\lambda}$) related to the Killing-Yano tensor.
Of course, there is no actual supersymmetry in the usual sense for the Kerr solution.

\section{NHEK=near-horizon limit of extreme Kerr}

Now we  turn to  the near-horizon region of the extreme Kerr black hole where life simplifies dramatically. Often in physics if we can not understand a system exactly, we first explore it in a limit where things simplify. In condensed matter physics the standard trick is to take the low-energy limit where atomic/molecular details can be ignored. In the present case of black holes we will go to small radii, but small radii correspond to high redshifts so this is also a low-energy limit. A heuristic explanation for why things simplify in the near-horizon limit is that
inside the ergosphere, physical objects have to rotate around with the black hole. At the horizon of the extreme black hole, they must rotate around at the speed of light and hence only chiral (co-rotating as opposed to counter-rotating) degrees of freedom appear. Purely chiral excitations are highly constrained. This is roughly why we expect quantum gravity to simplify in this limit.
\subsection{Extreme limit}
In the limit  $a=M$, formulae for the full Kerr geometry simplify:
 \begin{multline} \label{simp} \quad r_\pm=a=M,\quad S =2\pi M^2 =2\pi J, \\ T_H = 0,\quad \Omega_{H} = {a \over 2Mr_+}={1\over 2M}.~~~~~~~~~\end{multline}
 The metric reduces to
  \begin{multline} \label{ekerr} ds^2 = - {\Delta \over \rho^2}\left(d\hat{t} - a \sin^2\theta \, d\hat{\phi}\right)^2 +{\rho^2\over \Delta} \, d\hat{r}^2+\\
  {\sin^2\theta \over \rho^2}\left((\hat{r}^2+a^2)\, d\hat{\phi} -a\,d\hat{t}\right)^2 +\rho^2 d\theta^2,
 \end{multline}
 \beq\label{erhdl} \Delta = \left(\hat{r} -a\right)^2,\quad \rho^2 = \hat{r}^2 + a^2 \cos^2 \theta.\eeq
 \subsection{Near-horizon limit}
  We want to study the near-horizon region of the extreme Kerr solution. This region has enhanced isometries just like  the near-horizon AdS regions of various black brane solutions.     We  follow Bardeen and Horowitz \cite{BardeenPX} and zoom in on the region $r=M$ by introducing a scaling parameter $\lambda \rightarrow 0$. New coordinates
 \beq\label{scalr} r = {\hat{r}-M \over \lambda M }\eeq
 \beq\label{scalt} t = {\lambda \hat{t}\over 2M},\quad \phi = \hat{\phi} - {\hat{t}\over 2M}.\eeq
are held fixed as $\lambda$ is scaled. Note that for any finite $r$, the original radial coordinate $\hat r$ is forced to be very near the horizon value $M$ for  $\lambda \rightarrow 0$.
 In the limit, we obtain the smooth geometry
 \begin{multline}\label{nhek} ds^2 = 2\Omega^2 J \left[{dr^2 \over r^2} +d\theta^2 -r^2\, dt^2  \right.\\ \left.\phantom{{dr^2 \over r^2}}+ \Lambda^2(d\phi +r\,dt)^2\right],\end{multline}
involving two functions of $\theta$ given by
 \beq\label{oml} \Omega^2 = {1+\cos^2\theta\over 2}\, ,\quad \Lambda = {2\sin \theta \over 1+\cos^2\theta}\, .\eeq
 Formula (\ref{nhek}) is known as the near-horizon extremal Kerr geometry (NHEK). The black hole angular momentum $J$ appears only as an overall factor in front of the metric.
 Since this solution arises in the limit of a coordinate transformation of the Kerr solution, it is a solution to the Einstein equations.
 The metric (\ref{nhek}) is not asymptotically flat; in fact it has very peculiar asymptotics as $r\rightarrow \infty$.

 It is not hard to see that for slices of fixed polar angle $\theta$, we have a 3D  geometry which is locally a  ``warped'' version of $AdS_3$. At the special value $\theta=\theta_0:\Lambda(\theta_0)=1$ and the local metric is exactly that of $AdS_3$:
 \beq \label{adsl} ds^2 = 2\Omega^2 J \left[-r^2d t^2 +{dr^2\over r^2}+(d\phi+rdt)^2\right].\eeq
Since we know that gravity on AdS$_3$  always has a conformal symmetry \cite{BrownNW}, this is a strong  hint, which we follow up on below, of an underlying conformal symmetry  for extreme Kerr.

Globally, the  $\theta=\theta_0$ slice of NHEK is a quotient of $AdS_3$. The reason for this is that the angle $\phi$ is periodically identified
 $\phi\sim \phi+2\pi$.  We already know \cite{Maldacena} that a quotient of $AdS_3$ space is related to the finite temperature partition function of the dual conformal field theory.  Later we will see that the identification  $\phi\sim \phi+2\pi$ plays the same role in implying a finite  temperature in the context of Kerr/CFT .

The warped AdS$_3$ geometries appearing at generic fixed $\theta$ are of general interest in a number of contexts \cite{Nutku:1993eb,Gurses:1994,Rooman:1998xf,Bouchareb:2007yx,Herzog:2008wg,Anninos:2008qb,Anninos:2009zi,Anninos:2008fx,Compere:2008cv,Compere:2009zj,D'Hoker:2009bc,Detournay:2010rh} (for dS$_3$ cousins see \cite{Anninos:2009yc}). They are Lorentzian analogs of the squashed $S^3$. The ordinary round $S^3$ has the isometry group $SU(2)\times SU(2)$. Of course  $S^3$ is a Hopf fibration of $S^1$ over $S^2$  with
 a specific radius of the fiber. If we deform the radius of the $S^1$ fiber, we break  $SU(2)\times SU(2)$  down to $SU(2)\times U(1)$. Similarly, one can write quotients of $AdS_3$, which has isometry $SL(2,R)_R\times SL(2,R)_L$, as a Hopf-like fibration of $S^1$ over AdS$_2$ -- this is essentially equation (\ref{nhek}).  Warped AdS$_3$ is then obtained by deforming the fiber radius  and has the reduced  isometry group $SL(2,R)\times U(1)$.  Thus,  every section of fixed $\theta$ in the NHEK geometry has $SL(2,R)\times U(1)$.  This turns out to give the full isometry group of NHEK.

\section{Asymptotic Symmetry Group}

In order to make sense of  quantum gravity on the NHEK space we must specify boundary conditions. This is our
next problem. An immediate issue is that the boundary at $r\rightarrow \infty$ is rather peculiar. It does not look like the boundary of Minkowski space or that of AdS space. In order to define the theory we need to say something about the boundary conditions at this boundary.

 An important notion in doing this is the so-called asymptotic symmetry group (ASG), which is made up of the allowed diffeomorphisms modulo the trivial diffeomorphisms \beq\label{asg} \hbox{ASG} = {\hbox{Allowed diffeomorphisms}\over \hbox{Trivial diffeomorphisms} }.\eeq
 \subsection{Allowed diffeomorphisms}
To determine which are the allowed diffeomorphisms, we need to specify boundary conditions. Typically these will be of the form
\beq\label{bondcond} g_{ab}  = g_{ab}^{0} + \cO(r^{-p_{ab}}) \eeq
as $r$ approaches the boundary at infinity with $g^0_{ab}$ being a background metric and $p_{ab}$ a set of integers. Given the boundary conditions, the allowed diffeomorphisms $\xi$ are diffeomorphisms for which the variation  $\delta_\xi g$ of any allowed metric  $g$ results in a metric that is itself allowed.
 For example for 4D quantum gravity in asymptotic Minkowski space we typically demand that the metric approach the flat Minkowski  metric plus $1/r$ corrections. We then are not allowed to consider diffeomorphisms which go like
$r^{16}$ because this produces metric variations which do not vanish at infinity and therefore violate the boundary conditions. Note that the definition of the allowed diffeomorphisms does not require any information about dynamics.
\subsection{Trivial diffeomorphisms}
On the other hand, to know what the trivial diffeomorphisms are, we do need to know about the dynamics. Once dynamics are specified,
for every diffeomorphism $\zeta$ there is an associated charge $Q_\zeta$ canonically constructed to obey
\beq\label{charq} \{Q_\zeta,\Phi\}_{DB} = {\cal L}_\zeta \Phi \eeq
where $\{\cdot,\cdot\}_{DB}$ is the so-called Dirac bracket (the discussion at this point is classical) and $\Phi$ is any field in the theory. The Dirac bracket is a modification of the Poisson bracket designed to accommodate   local symmetries and the associated constraints. For example if we have a  discrete set of constraints $C_i$, then the Dirac bracket of two fields is defined as
\beq\label{dirbr} \{A,B\!\}_{DB} = \{A,B\!\} -\{A,C_i\!\}\{C_i,C_j\}^{-1} \{C_j,B\!\} \eeq
with $\{\cdot,\cdot\}$ being an ordinary Poisson bracket. This  bracket manifestly vanishes if $A$ is itself a constraint and so preserves the constraints. In gravity
the index $i$ is continuous and the inverse matrix $\{C_i,C_j\!\}^{-1}$ is a nonlocal Green's function.

For local symmetries the generator of the diffeomorphism $Q_\zeta$ has the generic form
\beq \label{gendf} Q_\zeta = \int\limits_{\hbox{boundary}} \hspace{-.3cm}X+ \int\limits_{\hbox{bulk}} C. \eeq
Typically the bulk term vanishes due to the constraint equations (Gauss' law for electromagnetism or the $G_{0\mu}$ constraint for gravity), so the
diffeomorphisms are always generated by boundary terms. The most familiar gravity example of this is the ADM mass. This generates time translations via the Dirac bracket and is the Hamiltonian of the theory. Because of the non-local nature of the Dirac bracket, it is possible for the boundary term by itself to generate a symmetry over the
entire spacetime.

Returning to our discussion of the ASG and the quotient (\ref{asg}), a trivial diffeomorphism  $\zeta$ is one for which $Q_{\zeta}=0$. For example any diffeomorphism that has compact support and vanishes at infinity is going to
be trivial.
The generators of the ASG can thus be thought of roughly as diffeomorphisms which are allowed by the boundary conditions but die off slowly enough at infinity to yield nonzero charges.
\subsection{Discussion}
In the quantum theory, the states should be annihilated by the trivial diffeomorphisms
and must transform in representations of the ASG.
Consequently this analysis provides us with the nontrivial symmetry group of the quantum theory.

There might appear to be much ambiguity in defining the theory because of the fact that we can start out with a large variety of boundary conditions. However, if we start with generic  boundary conditions we will run into trouble if they are either too strong or too weak. Suppose we start with 4D Minkowski space and demand that the metric components fall off as $\cO(1/ \sqrt{r})$ instead of the more usual $\cO(1/ r)$. Applying (\ref{charq}), we will find that the generator $Q$ is not always well-defined since the volume of the boundary goes like $r^2$, the integrand falls off like $r^{-3/2}$ (for some cases) and the integral (\ref{charq}) therefore diverges. This illustrates that  the generators do not exist if we impose boundary conditions which are too weak. In the case of too strong boundary conditions, for example $\cO( r^{-17})$, the energy is always  zero and the only consistent theory is one with a single state which is flat space with no gravitons or anything. So if we make the boundary conditions too weak, the generators are not well defined and if we make them too strong, the theory becomes trivial. In all the examples we know of there is only a very small window for interesting boundary conditions. Sometimes there is more than one consistent choice, but typically very few possibilities arise.

The asymptotic symmetry group is a very important and subtle concept. It may seem like we have described a fancy mechanism to reproduce results we already know, for example that the Poincar\'e group is the ASG of asymptotically spatially Minkowskian gravity. Indeed, it often happens that the ASG is simply the isometry group  of the vacuum. However this is not always the case. The most famous example is $AdS_3$ \cite{BrownNW}. For $AdS_4$, the ASG is the same as the vacuum isometry group $SO(3,2)$. Naively we might assume that similarly for $AdS_3$ the ASG is $SO(2,2)$, but in fact there is no consistent way to define a nontrivial theory in $AdS_3$ with
$SO(2,2)$ as the ASG.
If we want to accommodate any finite energy excitations we need  weaker boundary conditions - weak enough to give an ASG generated by  two copies of the Virasoro algebra. The resulting theory has  states transforming  in representations of the 2D  conformal group.
Simply by analyzing the ASG, we arrive at the far-reaching conclusion that any consistent definition of quantum gravity on $AdS_3$ space has to be a conformal theory in this sense. This was discovered by Brown and Henneaux in the 80's. Their analysis did not provide details about what the consistent definition, if any,  of quantum theory of gravity in $AdS_3$ should be.  Indeed it was a decade later \cite{StromingerSH} before we  found, using string theory and extra compact dimensions,  an  example of a consistent UV completion of quantum gravity in $AdS_3$ space.

Another highly nontrivial example is the ASG evaluated at future null infinity in asymptotically Minkowskian spaces. This is an infinite-dimensional group known as the BMS group. The full implications of this seem to be yet to be understood. Some interesting recent observations appear in \cite{Barnich}.

 \section{Kerr/CFT}
A lot of machinery had been developed over the years \cite{Compere:2008cv,Barnich:2001jy,Barnich:2003xg,Compere:2007az,Barnich:2007bf} to describe Dirac brackets and the $Q_\zeta$ generators. We now simply want to apply this machinery and find a consistent set of boundary conditions for NHEK. In our first analysis we  impose a restriction to ensure that our boundary conditions preserve extremality, i.e.
\beq\label{extr} \mathcal {E} = M^2-J=0.\eeq
We will discuss nonextremal cases later in these lectures. A motivation for studying extreme Kerr black holes first is that they are in some regards real-world analogs of the extreme black holes we have successfully analyzed in string theory, and may be easier to understand than the general case involving nonzero energy $\mathcal {E}$.  One set of consistent boundary conditions with $\mathcal {E}=0$ has been found. In the several years which have passed since the original Kerr/CFT paper \cite{GuicaMU} was published, nobody has found another solution to this problem. This suggests they may be unique but there is no proof.
\subsection{ASG for NHEK}
 The boundary conditions require
\begin{multline} \label{bound} h_{tt} = \cO(r^2),\quad h_{t\phi} = h_{\phi\phi} = \cO(1),\\
h_{\phi r}=h_{\theta\theta} = h_{\theta\phi} = h_{\theta t} = \cO(r^{-1}),\\
h_{rr} = \cO(r^{-3}),\quad h_{tr} = h_{\theta r} = \cO(r^{-2}). \end{multline}
In terms of the $h$ the $\mathcal {E}=0$ boundary condition is the vanishing of an ADM-like integral over the boundary.  The explicit expression is given in equation (\ref{hchang}) below with $\zeta=\p_\tau$.

The most general diffeomorphism preserving these boundary conditions is
\begin{multline} \label{aldiff} \zeta = \left[\e(\phi)+\cO(r^{-2})\right]\p_\phi +\left[- r \e'(\phi) +\cO(1)\right]\p_r\\
 +\left[C+\cO(r^{-3})\right]\p_\tau +\left[\cO(r^{-1})\right]\p_\theta \end{multline}
 where $\e(\phi)$ is an arbitrary smooth function and $C$ is an arbitrary constant. The subleading terms in (\ref{aldiff}) and leading $\p_\tau$ term correspond to the
 trivial diffeomorphisms i.e they do not contribute to $Q_\zeta$. Therefore we can take the leading part
 \beq\label{alrep}\zeta(\e )= \e(\phi)\p_\phi - r\e'(\phi)\p_r\eeq
as a representation of the quotient (\ref{asg}). In other words, this vector field generates the asymptotic symmetry group of the NHEK geometry.
 It is convenient to introduce a basis
\beq\label{zetan} \zeta_n \equiv \zeta(- e^{-in \phi}  )= -e^{-in\phi}\p_\phi -in\, e^{-in\phi} r\,\p_r.\eeq
The Lie bracket\footnote{$[X,Y] = {\cal L}_X Y - {\cal L}_Y X$ } of the vector fields $\zeta_n$ satisfies
\beq\label{aloop} i\left[\zeta_n,\zeta_m\right] = (m-n)\zeta_{m+n}.\eeq
From (\ref{aloop}) we learn that the asymptotic symmetry group is generated by one copy of the Virasoro algebra. There is no central charge here because this is just a classical vector field computation.

\subsection{Central charge}
Rather than studying the Lie brackets of $\zeta_n$, we now want to consider the Dirac bracket of the charges generating the $\zeta_n$  diffeomorphisms. By construction, the Dirac bracket algebra of the charges satisfies
\beq\label{charqq} \{Q_\zeta,\Phi\}_{DB} = {\cal L}_\zeta \Phi.\eeq
Using the Jacobi identity, this implies that
\beq\label{centvirr} \{Q_\zeta,Q_\xi\}_{DB} = Q_{[\zeta,\xi]}+c_{\zeta\xi}, \eeq
where $c_{\zeta\xi}$ is the central term. Amazingly enough, the central term in (\ref{centvirr}) can be calculated using classical gravity from the explicit expression for $Q_\zeta$. Infinitesimal charge differences between neighboring geometries $g_{\mu\nu}$ and $g_{\mu\nu}+h_{\mu\nu}$ are given by
\beq\label{hchang}\delta Q_{\zeta} = {1\over 8\pi G} \int_{\p \Sigma} K_{\zeta}(h,g) \eeq
where the integral is over the boundary of the spatial slice and
\begin{multline}\label{kzeta} K_\zeta (h,g) = -{1\over 4} \e_{\alpha\beta\mu\nu}dx^\alpha\wedge dx^\beta\Big[\zeta^\nu D^\mu h \\
- \zeta^\nu D_\sigma h^{\mu \sigma }+\zeta_\sigma D^\nu h^{\mu\sigma}+{1\over 2} hD^\nu \zeta^\mu  \\ +
{1\over 2}h^{\sigma \nu}(D^\mu \zeta_\sigma- D_\sigma \zeta^\mu )\Big].\end{multline} This expression and its many precursors have been studied in the literature for a long time, starting with ADM back in the 50's, and so has by now been worked out in great detail. From (\ref{hchang}) we can determine the central charge by applying it to fluctuations
$h = {\cal L}_\xi g$ ($g$ here is the NHEK metric)  produced by the allowed diffeomorphisms. This results in a very explicit expression for the central charge:
\beq\label{centr} c_{\xi\zeta} = {1\over 8\pi G} \int_{\p \Sigma} K_{\zeta}({\cal L}_\xi g,g).\eeq
In terms of the basis (\ref{zetan}) one finds
\beq\label{centrr}\{Q_{\zeta_m},Q_{\zeta_n}\}_{DB} = Q_{[\zeta_m,\zeta_n]} -iJ(m^3+2m)\delta_{m+n}.\eeq
The classical charges $Q_\zeta$ have units of action so we can use $\hbar $ to define a dimensionless quantity
\beq\label{gener}\hbar L_n = Q_{\zeta_n} +{3J\over 2}\delta_{n}\eeq
and then apply the usual quantization rules to the Dirac bracket $\{\cdot,\cdot \}_{DB} \rightarrow -{i \over \hbar }[\cdot,\cdot]$ for these. The quantum charge
algebra turns out to be
\beq\label{chalg}[L_m,L_n] = (m-n)L_{m+n} + {J\over \hbar}\, m(m^2-1)\delta_{m+n}.\eeq
From this we can read off the central charge for the NHEK geometry:
\beq\label{ccnhek} c = {12J \over \hbar},\eeq
where we have reinstated the factor of $\hbar$ normally set to one.
This means that the $\mathcal {E}=0$ states in NHEK with the boundary conditions (\ref{bound}) must form representations of one copy of the Virasoro algebra with central charge $ {12J \over \hbar}$  and hence in this sense comprise a (chiral half of) a 2D CFT\footnote{We henceforth take  $\hbar=1$.}.

From this analysis, we do not know if the CFT is local or unitary. We also cannot say, without an explicit microscopic construction, if the conformal symmetry persists beyond the semiclassical limit considered here.  However, since Kerr black holes do exist, we do know that the structure we have found arises as a limit of a consistent physical system.

While 2D conformal symmetry often appears in critical point behavior of real-world condensed matter systems, this is the first time it has arisen in limits of observable astronomical systems.
The 2D conformal symmetry arises here with a holographic connection to
the 4D diffeomorphism group (note the radial $\epsilon'\p_r$ term in (\ref{alrep})).  This connection opens the door to adapting the beautiful insights in string theory holography to real-world black holes.

\subsection{Cardy entropy}
Conformal field theories have many universal properties and we would like to map these to the many universal properties of black hole dynamics. The first thing we would like to understand is whether or not there is some way for us to recover the Bekenstein - Hawking formula for black hole entropy. For the extreme Kerr geometry $S_{BH}=2\pi J$.  One would like to understand this in terms of the microstate degeneracy of a CFT at finite temperature. This is at first  puzzling since we know that extreme Kerr has zero Hawking temperature. However there are two relevant conserved quantities. One of these is the energy which is conjugate to the Hawking temperature and the other is an angular momentum conjugate to the angular velocity. We are interested in counting extreme Kerr microstates which are moving at the speed of light in the near-horizon region of the black hole. Roughly speaking this means that the quantum number we are after is a linear combination of the energy and the angular momentum. Let's be specific.

In general, a black hole represents not a pure, but a mixed, state and the quantum fields around the black hole are in a thermal state. For a Schwarzschild
black hole, the thermal state is weighted by the Boltzmann factor
\beq\label{thstate} e^{-\omega / T_H}.\eeq
Adding angular momentum changes this to
\beq\label{athstate} e^{-(\omega - m \Omega_H) / T_H}\eeq
with $\Omega_H$ being the angular velocity of the horizon which for extremal Kerr is $\Omega_H = 1/ 2M$. When we carefully take the extreme limit of Kerr,
$T_H$ goes  to zero, but this does not necessarily mean that the weighting factor (\ref{athstate}) is trivial. The reason for this is the possibility of states which have $\omega$ very near $m \Omega_H$. In the extreme limit $T_H\to 0$, we get $\omega=m \Omega_H$ states contributing to a density matrix weighted by the angular momentum $m$. Carefully taking the limit one finds for ${T_H\to 0}$
\beq\label{mixstate} e^{-(\omega - m \Omega_H) / T_H} \rightarrow e^{-2\pi m} = e^{-m/ T_L}.\eeq
From here we can read off the relevant temperature as $T_L = 1/ 2\pi$. We can also note that the angular momentum $m$ is the eigenvalue of $L_0$. Using the Cardy entropy formula for a chiral CFT, we find the entropy obeys
\beq\label{mcardy} S_{cardy} = {\pi^2\over 3} c\, T_L = {\pi^2\over 3} 12 J\, T_L  = 2\pi \! J.\eeq
which  matches the Bekenstein - Hawking entropy (\ref{tentr}) exactly!

 Note that since we do not know the exact CFT, we cannot be sure that
it is in the class of unitary and modular invariant theories for which the Cardy formula is known to apply. We only know some very general properties of this CFT. In order to provide an exact description we would need to know all the microscopic laws of physics to the Planck scale and beyond. In string theory we take a top-down approach: we assume some microscopic description and that way we know everything about the dual CFT.  That is not what we are doing in these lectures. Rather we are taking a bottom-up approach and must live with some uncertainty about the details of the CFT.  We are trying to understand how the universal properties of black holes might follow from any consistent UV completion of the known laws of physics.  We are not trying to find the completion. So rather than try to prove that the
Cardy formula is applicable, we take the precise agreement of the Bekenstein-Hawking and Cardy entropy formulas resulting from our ASG analysis as the first piece of evidence both for the applicability of the Cardy formula and the physical  relevance of the picture of extreme Kerr as a 2D CFT. More evidence will appear below.

Roughly speaking, the Cardy formula works when
we are justified in applying statistical reasoning.
It pertains to the statistical  limit of 2D CFTs and hence is related to the
statistical limit of black holes as studied in  \cite{Preskill}.
A sufficient, though not necessary, condition for the Cardy formula to be valid is that the temperature is large compared to the central charge. Clearly this is not the case here. It was also not the case for the stringy black holes studied in  \cite{StromingerSH}. There, the temperature was just one of the charges and in the limit when the supergravity approximation was valid, the temperature was parametrically small compared to the central charge.
Often, a sufficient condition for the applicability of the Cardy formula is that the temperature is large relative to the gap or lightest excitation of the theory, which means that a large number of degrees of freedom are excited. Essentially it is just a thermodynamic approximation applied to the many degrees of freedom in the CFT. We actually know that the
gap for extreme Kerr must be very small. It was computed in a paper by Preskill, Schwarz, Shapere, Trivedi and Wilczek \cite{Preskill}  two decades ago using only semiclassical reasoning. They showed that if the gap for extreme black holes is sufficiently large, the Hawking calculation would be invalid since a black hole can not radiate quanta with energies less than the gap. Therefore a large gap is inconsistent with the semiclassical Hawking calculation. This argument can provide also us with an upper bound on the size of the gap. For case of Kerr, the $L_0$ gap is of order
$1/ J$. The temperature $T_L={1 \over 2 \pi}$ is already much larger than this, so the Cardy formula may plausibly apply.  A similar story justified  the string theory computations \cite{jmls}.

So far in our matching between gravity and the CFT, we have found agreement with just one number. However there are many generalizations of this construction in which the central charge and entropy are complicated functions of various geometrical data and the functions are matched in entirety \cite{Lu:2008jk,Azeyanagi:2008kb,Hartman:2008pb,Chow:2008dp,Isono:2008kx,Azeyanagi:2008dk,Peng:2009ty,Chen:2009xja,Loran:2009cr,Ghezelbash:2009gf,Lu:2009gj,Compere:2009dp,Astefanesei:2009sh,Garousi:2009zx,Azeyanagi:2009wf,Wu:2009di,Becker:2010jj,Becker:2010dm}. Further, and qualitatively different, supporting evidence for Kerr/CFT comes from scattering amplitudes in NHEK. Kerr scattering amplitudes were computed long ago in \cite{StarobAP,StarobCH,TeukolSK,TeukolPR,TeukolPRT}. They turn out to be complicated products involving ratios of  four gamma functions with arguments depending on the energies and spins of the black hole and the incoming and outgoing waves . It has been shown in a variety of contexts  \cite{BredbergPV,Cvetic:2009jn,Hartman:2009nz,Chen:2010ni} that
these correspond exactly to the finite temperature CFT correlators! Hence the picture of extreme Kerr as a CFT holds together well.
\subsection{Beyond extremality}
The natural next question is how to go beyond extremality, i.e. to the case where $\mathcal {E} = M^2-J >0$. For infinitesimally small deviations from extremality, Castro and Larsen \cite{CastroJF} and others \cite{Matsuo:2009sj,Matsuo:2009pg,Rasmussen:2009ix} found an alternate set of boundary conditions which lead to a second copy of the Virasoro algebra. Since the conserved charge $\mathcal {E}$ derives from an $SL(2,R)$ isometry of NHEK, the relevant Virasoro algebra is an enhancement of the $SL(2,R)$ isometry rather than the enhancement of the $U(1)$ described above.

In order to go beyond linear deviations from extremality, we need to include the effects of backreaction. The backreacted geometry is no longer asymptotically NHEK and the whole picture breaks down.   It is not clear exactly what this means or how to proceed further at this point. One approach is to try to get some insight by embedding Kerr/CFT in string theory and see how the issue resolves itself in that context.\footnote{  Some recent progress along those lines appeared, after these lectures were given, in  \cite{Guica:2010ej,Compere:2010uk}.}

We will not further explore this direction in these lectures. However in the next lecture we will find encouraging results by approaching the issue from a different angle. We simply analyze the form of the scattering amplitudes and entropy functions and find surprisingly that at low energies they take exactly the form expected for a CFT --  even far from extremality! We have no a priori explanation for this ``hidden'' conformal symmetry but it seems to indicate  more Kerr miracles still await discovery.

 \section{Hidden conformal symmetry for generic Kerr }
 \subsection{Wave equation as $SL(2,R)$ Casimir}
 Consider the scalar Laplacian $\nabla^2$ in the Kerr geometry. Using the Killing-Yano tensor we can construct a second-order differential operator ${f_{\alpha}}^\mu f^{\alpha \nu }\nabla_\mu \nabla_\nu$ which commutes with $\nabla^2$. This allows the wave equation to be separated  \cite{Carte}:
 \beq\label{separ} \Phi(t,r,\theta,\phi) = e^{-i\omega t +i m \phi} R(r) S(\theta).\eeq
Denoting the separation constant $K_\ell$ we have
  \begin{multline}\label{lapl} -K_l S(\theta)=\left[{1 \over \sin \theta} \p_\theta (\sin \theta \p_\theta) \right.\\\left.- {m^2 \over \sin^2 \theta }  +\omega^2 a^2 \cos^2 \theta \right]S(\theta), \end{multline}
  and
 \begin{multline}\label{lapll}K_l R(r)=\left[\p_r(\Delta \p_r) + {(2Mr_+\omega - am)^2\over (r-r_+)(r_+-r_-)}\right.\\ -{(2Mr_-\omega - am)^2\over (r-r_-)(r_+-r_-)}\\ \left. \phantom{{(2Mr_-)^2\over(r_+)}}+(r^2+2M(r+2M))\omega^2\right]R(r).\end{multline}
 Both of these equations are solved by Heun functions which are known only numerically. However, we can try to simplify the equations by going to low frequencies. The last term in (\ref{lapl}) can be neglected if $\omega M \ll 1$ i.e. for excitation wavelengths larger than the black hole radius. In this case, the geometry can be divided into
 two regions
  \begin{align}\label{region} \hbox{``Near''} &: \quad r\ll {1\over \omega},\notag\\
  \hbox{``Far''} &: \quad r\gg M \end{align}
  with large overlaps.
We can then solve the equations (\ref{lapl}) and (\ref{lapll}) in both the near and far regions, and match the solutions at any value $r_{match}$ of $r$  in the  intermediate region
 \beq\label{match} M\ll r_{match}\ll {1\over \omega}.\eeq
 The fact that, in this approximation, the answer cannot depend on  $r_{match}$ implies that the solutions in the near region must behave in some special symmetric way under transformations of $r_{match}$. Moreover since the redshift factor depends on $r$, transformations of $r_{match}$ are tied to scale transformations. This strongly hints that the near region physics should exhibit some kind of conformal symmetry arising from invariance of the full answer under shifts of the matching surface. We will see below that this is indeed the case.

In the limit $\omega M\ll 1$, the equation (\ref{lapl}) simplifies to the spherical Laplacian and is solved in terms of spherical harmonic functions with $K_l = l(l+1)$. The radial equation (\ref{lapll}) in the far region is solved by Bessel functions. In the near region, the equation simplifies to
 \begin{multline}\label{slapll} \left[\p_r(\Delta \p_r) + {(2Mr_+\omega - am)^2\over (r-r_+)(r_+-r_-)}\right.\\ \left.-{(2Mr_-\omega - am)^2\over (r-r_-)(r_+-r_-)}\right]R(r)=l(l+1) \, R(r),\end{multline}
 whose solutions are hypergeometric functions. In order to obtain the solution for the full geometry we need to do a matching between the two regions. The fact that
 the hypergeometric functions form representations of $SL(2,R)$ gives us further reason to expect some hidden conformal symmetry.

 This near region $SL(2,R)$ symmetry that appears in the Laplace equation for scalar perturbations can be made explicit by introducing natural `conformal' variables
 \begin{align}\label{confw} w^+ &= \sqrt{r-r_+ \over r-r_-}\, e^{2\pi T_R \phi}\notag\\
  w^- &= \sqrt{r-r_+ \over r-r_-}\, e^{2\pi T_L \phi - t/2M} \\
  y  &= \sqrt{r-r_+ \over r-r_-}\, e^{\pi (T_R+T_L) \phi - t/4M} \notag\end{align}
 where
 \beq\label{tlr} T_L= {r_++r_- \over 4\pi a},\quad T_R = {r_+-r_- \over 4\pi a}.\eeq
 We can define the vector fields
 \begin{align}\label{vect} H_1 &= i\p_+ \notag\\
 H_0 &= i\left(w^+ \p_+ +{1\over 2} y\p_y \right)\\
 H_{-1} &= i\left(w^{+2}\p_+ +w^+ y\p_y - y^2 \p_-\right) \notag\end{align}
 and
  \begin{align}\label{vectt} \bar{H}_1 &= i\p_- \notag\\
 \bar{H}_0 &= i\left(w^- \p_- +{1\over 2} y\p_y \right)\\
 \bar{H}_{-1} &= i\left(w^{-2}\p_- +w^- y\p_y - y^2 \p_+\right) \notag\end{align}
 which obey the $SL(2,R)$ algebra
 \beq\label{sla}[H_0,H_{\pm 1}] = \mp i H_{\pm1},\quad [H_{-1},H_{1}]=-2iH_0 ,\eeq
and similarly for $\bar{H}_0$ and $\bar{H}_{\pm1}$.
 The $SL(2,R)$ quadratic Casimir, written in  $(t,r,\theta,\phi)$  variables, is
 \begin{multline}\label{casim}  \mathcal {H}^2  = \bar{\cal H}^2= \p_r(\Delta \p_r)\, +\\ {(2Mr_+\omega - am)^2\over (r-r_+)(r_+-r_-)}-{(2Mr_-\omega - am)^2\over (r-r_-)(r_+-r_-) }\, .\end{multline}
 As a result of this, the equation (\ref{slapll})  for $\Phi(r)$ in the near region can be written
 \beq\label{clapl} \mathcal {H}^2 \Phi = \bar{\cal H}^2 \Phi = l(l+1) \Phi.\eeq
This gives the solutions for $\Phi$ as representations of $SL(2,R)$ with conformal weights
 \beq\label{confwb} (h_L,h_R) = (l,l) .\eeq
However, although $SL(2,R)_L \times SL(2,R)_R$ is a symmetry of the near region scalar wave equation, it is broken by
the extra periodic identification
 \beq\label{ident} \phi \sim \phi +2\pi.\eeq
Under the identification (\ref{ident}), the conformal coordinates identify as
 \be\label{cident} w^+\sim  e^{4\pi^2 T_R}w^+, \ w^-\sim e^{4\pi^2 T_L} w^-,\\ y\sim  e^{2\pi^2 (T_R+T_L)}y .\ee
This identification is generated by
 \begin{multline}\label{elem} g = e^{-4\pi^2i T_R H_0 - 4\pi^2 i T_L \bar{H}_0},\\g \in SL(2,R)_L\times SL(2,R)_R,\end{multline}
which is exactly the form of the identification for  a CFT partition function at finite temperature $(T_L, T_R)$.

  We stress that the near region discussed above is not a ``near-horizon'' region since $r$ can be arbitrarily large for fixed $M$ and the conformal symmetry, although it acts on solutions of the wave equation, is not a geometric isometry!

\subsection{Cardy entropy}
Now let us simply suppose there is some underlying CFT at the temperatures (\ref{tlr}) for generic Kerr. Its entropy would be given by  the Cardy formula
 \beq\label{cardy} S  = {\pi^2\over 3} \, c_LT_L +{\pi^2\over 3} \, c_R T_R.\eeq
We only know the central charges $c_L=c_R=12 J$ for the extreme case and first order variations away from extremality. So we further assume that the central charges behave smoothly and that they do not change as we move away from the extremal case. Using (\ref{tlr}) for the temperatures, we arrive at
 \begin{multline}\label{micrent} S_{Cardy} = {\pi^2 \over 3} 12 J \left({r_++r_-\over 4\pi a} + {r_+-r_- \over 4\pi a}\right) \\= {2\pi r_+ J \over a} = 2\pi Mr_+=S_{BH}!\end{multline}
Comparing this to the  Bekenstein - Hawking formula (\ref{tentr}), we see an impressive and detailed match for general $M,J$ Kerr black holes. One can also look at the near-region low-energy scattering amplitudes. These agree precisely with those of a CFT  \cite{CastroFD} even for nonextremal Kerr.
Hence the scattering amplitudes exhibit symmetries which   do not come from manifest symmetries of the action.
Another example of such `hidden' symmetries which are not manifest in the action but appear in the scattering amplitudes  is described in lectures at this school by N. Arkani-Hamed on   4D ${\cal N}=4$
SYM theory.

Let us summarize. There is no a priori ASG or other type of argument indicating that black holes with generic values of $M$ and $J$ are dual to 2D CFTs. Nevertheless a phenomenological analysis of the scattering amplitudes and entropy formulae reveals that they exhibit a hidden  2D conformal symmetry:  yet another `miracle' of the Kerr story.  The challenge now is to understand why this should be so.

\section{Future directions}

While some things have been understood at this point,  there are several remaining, interrelated puzzles/questions concerning the theoretical structure of Kerr/CFT, some of which we have mentioned along the way.  Related quantum gravity questions were explored in the previous Carg\`ese lectures \cite{StromingerAJ}.
These Kerr/CFT puzzles/questions include

\noindent{\bf(i)}Boundary conditions are known which give either a left or a right -moving Virasoro generating the ASG. Are there boundary conditions which simultaneously give both? If not, why not?

\noindent{\bf (ii)}How do we understand the fact that at finite $M^2-J$ the NHEK region disappears?

\noindent{\bf (iii)}How do we match the instability of extreme Kerr due to quantum Unruh-Starobinsky superradiance to properties of the CFT?

\noindent{\bf (iv)}Where did the hidden conformal symmetry come from and why does the Cardy formula work?
Is there a generalization of the standard ASG analysis which can be applied to the $r\ll { 1 \over \omega}$ near-region to explain the hidden conformal symmetry?

\noindent{\bf (v) } One might hope to match a thermal state in a nonchiral CFT
with the general Kerr black hole. Such a state  has three parameters: $c$,  $T_L$ and $T_R$, while Kerr has only $M$ and $J$, and can therefore at best describe a subspace of the CFT states. Where does the restriction to a subspace come from? Is there a holographic dual for the generic CFT state?

Perhaps the most promising approach to addressing these questions is by embedding Kerr/CFT into string theory. Indeed many aspects of the parallel Brown-Henneaux analysis of AdS$_3$ were not understood until a stringy embedding was discovered.  Recently (after these lectures were given), progress has been made in this direction in the context of charged spinning black holes
in five dimensions with a Kaluza-Klein $S^1$ \cite{Guica:2010ej}. The structure of the corresponding NHEK geometry depends on the value of the electric charge.
When the charge is pushed to its maximal value, the NHEK  geometry locally approaches $AdS_3 \times S^3$. The dual at the maximality is then easily  identified with standard stringy methods and has the central charge predicted by Kerr/CFT. The linearized properties  away from this maximal point also agree with Kerr/CFT expectations. Even more recently \cite{Wei}, the
dual theory has been identified  for all finite submaximal values of the charge as the low energy limit of a certain wrapped D-brane configuration with nonzero charge densities, and found to be related to the so-called dipole-deformed gauge theories. One hopes that this intricate construction can be used to shed light on the general puzzles discussed above, and suggest answers which do not depend on the details of string theory.

Of course since Kerr/CFT applies to the real world, it is not necessary to understand all aspects of its theoretical structure before attempting to find observational consequences of the symmetry. It is often the case in physics that the comparison of theory to experiment is an ingredient in understanding both!  Kerr/CFT is a statement about a symmetry which applies to objects seen  in the sky.  At present, the observational data from those objects is  good and rapidly improving. There are many observed phenomena which are not understood \cite{McClintock:2006xd,McClintock:2011zq}. Symmetries often provide both predictions about and a useful framework for organizing the observational data. They also suggest what to look for. Analyses of black hole properties tend to focus either on the non-rotating case or small perturbations in the rotation parameter. Perhaps interesting simplifications and structures occur near extremality. This direction seems ripe for  exploration in the near future.

\noindent

 \centerline{\bf Acknowledgements}

This article was transcribed from tapes of lectures given by A. Strominger at the 2010 Carg\`ese School on String Theory.
We are grateful to L. Baulieu, J. de Boer, M. Douglas, E. Rabinovici, P. Vanhove and P. Windey for the invitation to lecture and for organizing an exceptionally stimulating school. We would also like to thank the participants of the school for many stimulating conversations and questions.

\end{document}